\documentclass[aps,prl,showpacs,floatfix,twocolumn,superscriptaddress,longbibliography]{revtex4-2}

\usepackage{float}
\usepackage{color}
\usepackage{bm}
\usepackage{hyperref}
\usepackage{todonotes}
\usepackage{verbatim}
\usepackage{soul}
\usepackage{glossaries}
\usepackage{sidecap}
\usepackage{hyperref}
\hypersetup{
    colorlinks=true,
    linkcolor=blue,
    filecolor=magenta,      
    urlcolor=red,
    citecolor=blue,
}

\def\Q{\ensuremath{\bm{Q}}}
\def\TN{\ensuremath{T_\mathrm{N}}}
\def\Ba4310{Ba$_4$Ir$_3$O$_{10}$}
\def\Sr4310{(Ba$_{1-x}$Sr$_x$)$_4$Ir$_3$O$_{10}$}
\def\Sp{$|S=1/2\rangle^{+}$}
\def\Sm{$|S=1/2\rangle^{-}$}

\newacronym{RIXS}{RIXS}{resonant inelastic x-ray scattering}
\newacronym{REXS}{REXS}{resonant elastic x-ray scattering}
\newacronym{FWHM}{FWHM}{full-width at half-maximum}
\newacronym{AFM}{AFM}{antiferromagnetic}
\newacronym{SOC}{SOC}{spin-orbit coupling}
\newacronym{1D}{1D}{one-dimensional}
\newacronym{2D}{2D}{two-dimensional}
\newacronym{3D}{3D}{three-dimensional}
\newacronym{NN}{NN}{nearest neighbor}
\newacronym{QSL}{QSL}{quantum spin liquid}
\newacronym{ED}{ED}{exact diagonalization}
\newacronym{XAS}{XAS}{x-ray absorption spectrum}
\newacronym{INS}{INS}{inelastic neutron scattering}
\newacronym{DFT}{DFT}{density functional theory}
\newacronym{GGA}{GGA}{generalized gradient approximation}
\newacronym{CEF}{CEF}{crystal electric field}
\newacronym{TM}{TM}{transition-metal}
\newacronym{DMFT}{DMFT}{dynamical mean field theory}

\begin{document}

\title{Emergence of spinons in layered trimer iridate Ba$_4$Ir$_3$O$_{10}$}

\author{Y. Shen}\email[]{yshen@bnl.gov}
\affiliation{Condensed Matter Physics and Materials Science Department, Brookhaven National Laboratory, Upton, New York 11973, USA}

\author{J. Sears}
\affiliation{Condensed Matter Physics and Materials Science Department, Brookhaven National Laboratory, Upton, New York 11973, USA}

\author{G. Fabbris}
\affiliation{Advanced Photon Source, Argonne National Laboratory, Argonne, Illinois 60439, USA}

\author{A. Weichselbaum}
\author{W. Yin}
\affiliation{Condensed Matter Physics and Materials Science Department, Brookhaven National Laboratory, Upton, New York 11973, USA}

\author{H. Zhao}
\affiliation{Department of Physics, University of Colorado Boulder, Boulder, Colorado 80309, USA}

\author{D. G. Mazzone}
\affiliation{Laboratory for Neutron Scattering and Imaging, Paul Scherrer Institut, CH-5232 Villigen, Switzerland}

\author{H. Miao}
\affiliation{Condensed Matter Physics and Materials Science Department, Brookhaven National Laboratory, Upton, New York 11973, USA}
\affiliation{Material Science and Technology Division, Oak Ridge National Laboratory, Oak Ridge, Tennessee 37830, USA}

\author{M .H. Upton}
\author{D. Casa}
\affiliation{Advanced Photon Source, Argonne National Laboratory, Argonne, Illinois 60439, USA}

\author{R. Acevedo-Esteves}
\author{C. Nelson}
\author{A. M. Barbour}
\author{C. Mazzoli}
\affiliation{National Synchrotron Light Source II, Brookhaven National Laboratory, Upton, New York 11973, USA}

\author{G. Cao}
\affiliation{Department of Physics, University of Colorado Boulder, Boulder, Colorado 80309, USA}

\author{M. P. M. Dean}\email[]{mdean@bnl.gov}
\affiliation{Condensed Matter Physics and Materials Science Department, Brookhaven National Laboratory, Upton, New York 11973, USA}

\date{\today}

\begin{abstract}
Spinons are well-known as the elementary excitations of one-dimensional antiferromagnetic chains, but means to realize spinons in higher dimensions is the subject of intense research. Here, we use resonant x-ray scattering to study the layered trimer iridate \Ba4310{}, which shows no magnetic order down to 0.2~K. An emergent one-dimensional spinon continuum is observed that can be well-described by XXZ spin-1/2 chains with magnetic exchange of $\sim$55~meV and a small Ising-like anisotropy. With 2\% isovalent Sr doping, magnetic order appears below \TN{}=130~K along with sharper excitations in \Sr4310{}. Combining our data with exact diagonalization calculations, we find that the frustrated intra-trimer interactions effectively reduce the system into decoupled spin chains, the subtle balance of which can be easily tipped by perturbations such as chemical doping. Our results put \Ba4310{} between the one-dimensional chain and two-dimensional quantum spin liquid scenarios, illustrating a new way to suppress magnetic order and realize fractional spinons.
\end{abstract}

\maketitle


\Glspl*{QSL} are novel states of matter where quantum fluctuations prevent symmetry breaking down to zero temperature \cite{Anderson1973Resonating,Anderson1987Resonating,Balents2010Spin,Lucile2017Quantum,Zhou2017Quantum,Broholm2020quantum}. A key characteristic of \glspl*{QSL} is that they can host fractional elementary excitations called spinons, carrying spin-1/2, which can serve as a fingerprint for these states. Arguably the best-understood example of this is a \gls*{1D} spin-1/2 \gls*{AFM} chain where the spinons describe the spin domain wall dynamics \cite{Lake2005Quantum,Mourigal2013Fractional,Bera2017Spinon, Faure2018Topological, Faure2019Tomonaga,Schlappa2012spin,Bisogni2015Orbital,Kumar2021Unraveling,Rossi2020Jeff}. \Glspl*{QSL} in higher dimensions are harder to identify, but hold promise for realizing novel topological order and intrinsic long-range quantum entanglement with potential applications in quantum information \cite{Wen2002Quantum, Ioffe2002Topologically}. Several studies have provided evidence for spinons in \gls*{2D} or \gls*{3D} systems such as triangular, Kagome, Kitaev honeycomb, or pyrochlore lattices \cite{Han2012Fractionalized,Shen2016Evidence,Gao2019Experimental,Martinelli2021Fractional,Scheie2021Witnessing,Banerjee2017Neutron}, while definitive material-realization remains controversial and is a major target of research. Another possible approach relies on \gls*{1D} systems as building blocks to realize higher-dimensional \gls*{QSL} states \cite{Kohno2007Spinons}, which, however, is less explored.

\begin{figure}
\includegraphics{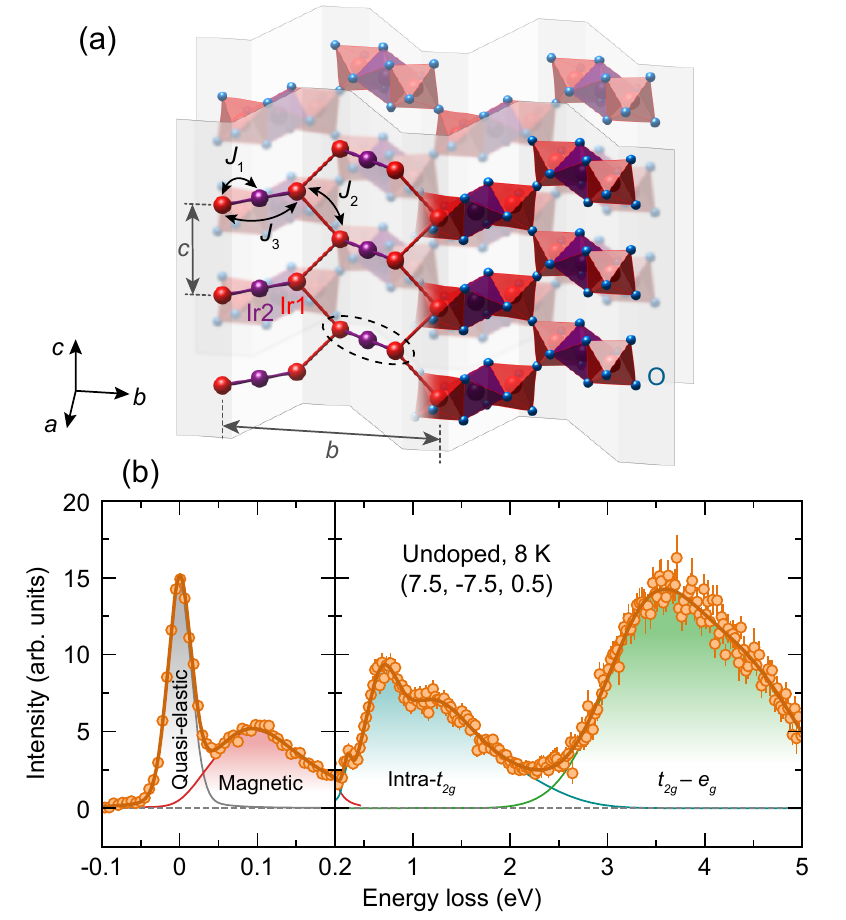}
\caption{(a) Crystal structure of \Ba4310{}. The material contains two symmetry-inequivalent Ir ions: Ir1 and Ir2, occupying the outer and middle sites of the trimers (black dashed ellipse), respectively. $J_n$ where $n=1,2,3$ denotes magnetic interactions for the first, second, and third nearest Ir neighbors. The barium ions are not shown. (b) Representative \gls*{RIXS} spectrum of \Ba4310{} at the Ir $L_3$-edge with a constant background subtracted. The circles represent the data and the different lines with shaded areas indicate the fitting results of different components, the summation of which gives the dark orange curve. Error bars represent 1 standard deviation based on Poisson statistics.}
\label{fig:structure}
\end{figure}


The insulating magnet \Ba4310{} is an intriguing candidate for realizing novel mechanisms for the emergence of spinons \cite{Cao2020Quantum,Cao2020Quest,Chen2021Structural,Sokolik2021Spinons}. As shown in Fig.~\ref{fig:structure}(a), it has a quasi-\gls*{2D} structure composed of buckled sheets with a Ir $5d^5$ nominal atomic configuration. Each sheet constitutes corner-connected Ir$_3$O$_{12}$ trimers containing three distorted face-sharing IrO$_6$ octahedra. The shortest Ir-Ir bond is the one between Ir1 and Ir2 within a trimer, which has a length of $\sim$2.58~\AA{} \cite{Stitzer2002Crystal,Cao2020Quantum}, shorter than that of elemental iridium ($\sim$2.71~\AA{} \cite{Wyckoff1963Crystal}). Thus, strong intra-trimer couplings are expected, such as those found in other face-sharing iridates \cite{Terasaki2016Novel,Okazaki2018Spectroscopic,Ye2018Covalency,Wang2019Direct}, and iso-structural Ba$_4$Ru$_3$O$_{10}$ \cite{Klein2011Antiferromagnetic,Streltsov2012Unconventional,Igarashi2013Xray,Radtke2013Magnetism,Igarashi2015Effects,Sannigrahi2021Orbital}. This directly leads to a strong \gls*{NN} exchange interaction $J_1$ within each trimer [see Fig.~\ref{fig:structure}(a)]. Another intra-trimer interaction is the third \gls*{NN} term $J_3$, which can be realized by the superexchange path through the Ir2 ion. The Ir1 ions are also connected through the second \gls*{NN} interaction $J_2$, forming zig-zag chains along the crystalline $c$ direction. All other magnetic interactions can be ignored due to the long bond lengths and unfavorable hopping trajectories. Considering the expectation of appreciable magnetic exchange and the lack of any obvious magnetic frustration, this material would be expected to be a long-range ordered antiferromagnet. Previous transport and magnetization studies found no magnetic order down to 0.2~K despite a Curie-Weiss temperature up to -766 K \cite{Cao2020Quantum}. In contrast, the material shows a linear behavior in the low-temperature magnetic heat capacity \cite{Cao2020Quantum}, resembling a gapless \gls*{QSL} \cite{Yamashita2008Thermodynamic,Yamashita2011Gapless}. In fact, the ground state is rather susceptible to perturbations including sample growth conditions and chemical doping \cite{Cao2020Quantum,Cao2020Quest,Chen2021Structural}. The origin of the highly suppressed magnetic order and fragile ground state remains puzzling and cannot be solved without direct measurements of the magnetic excitation spectrum.

In this paper, we use \acrfull*{RIXS} at the Ir $L_3$-edge and \gls*{REXS} at the O $K$-edge to study the magnetic properties of  \Ba4310{}. For comparison, we also measure the isovalently doped \Sr4310{} ($x$=0.02), in which a small amount of Sr doping surprisingly triggers magnetic order below 130~K. \gls*{1D} continuous spinon excitations are discovered in undoped \Ba4310{} which can be well-described by a spin-1/2 XXZ \gls*{AFM} chain with small Ising-like anisotropy. In contrast, \Sr4310{} shows magnetic order with propagation vector of (0.5, 0, 0), and a comparably sharp dispersion, which remains \gls*{1D} like down to base temperature. Through \gls*{ED} calculations, we show that the zig-zag chains retained by $J_2$ are effectively decoupled at a critical point triggered by the competition between the inter-chain interactions $J_1$ and $J_3$, the balance of which can be easily tipped by perturbations.


\gls*{RIXS} data were collected at the Ir $L_3$-edge with a horizontal scattering plane and $\pi$ polarization and energy resolution of about 32~meV \cite{supp}. Wavevectors throughout the manuscript are defined using standard reciprocal lattice units (r.l.u.) notation as \Q{}$=H\mathbf{a}^{\ast}+K\mathbf{b}^{\ast}+L\mathbf{c}^{\ast}$ based on lattice constants $a=$7.2545~\AA, $b=$13.192~\AA, $c=$5.7737~\AA, $\alpha=\gamma=90^{\circ}$, $\beta=113.513^{\circ}$.

\begin{figure}
\includegraphics{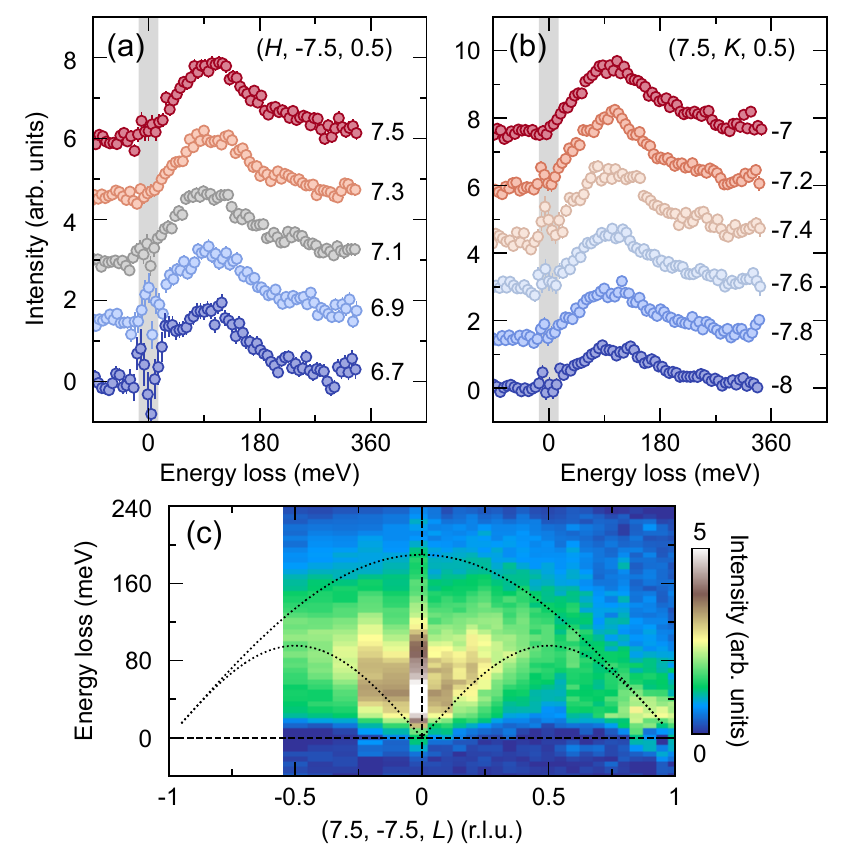}
\caption{One-dimensional spinons in undoped \Ba4310{} at 8~K. (a),~(b) Magnetic excitations at different \Q{} along $H$ and $K$ directions, respectively, showing essentially dispersionless behavior. All the \gls*{RIXS} spectra in the text are presented with the constant background, quasi-elastic line and high-energy $dd$ excitations subtracted to highlight the magnetic contributions \cite{supp}. The vertical gray bars indicate the quasi-elastic regime, the widths of which represent the energy resolution. The values of $H/K$ are indicated for each curve and the spectra are shifted along the $y$ axis for clarity. (c) Colorplot of the magnetic excitations along the $L$ direction. The dotted lines are the calculated excitation boundaries of an \gls*{AFM} spin-1/2 chain with $J_\text{chain}=55$~meV and $\Delta=1.3$. The dashed lines are guides to the eye.}
\label{fig:undoped}
\end{figure}


\begin{figure*}
\includegraphics{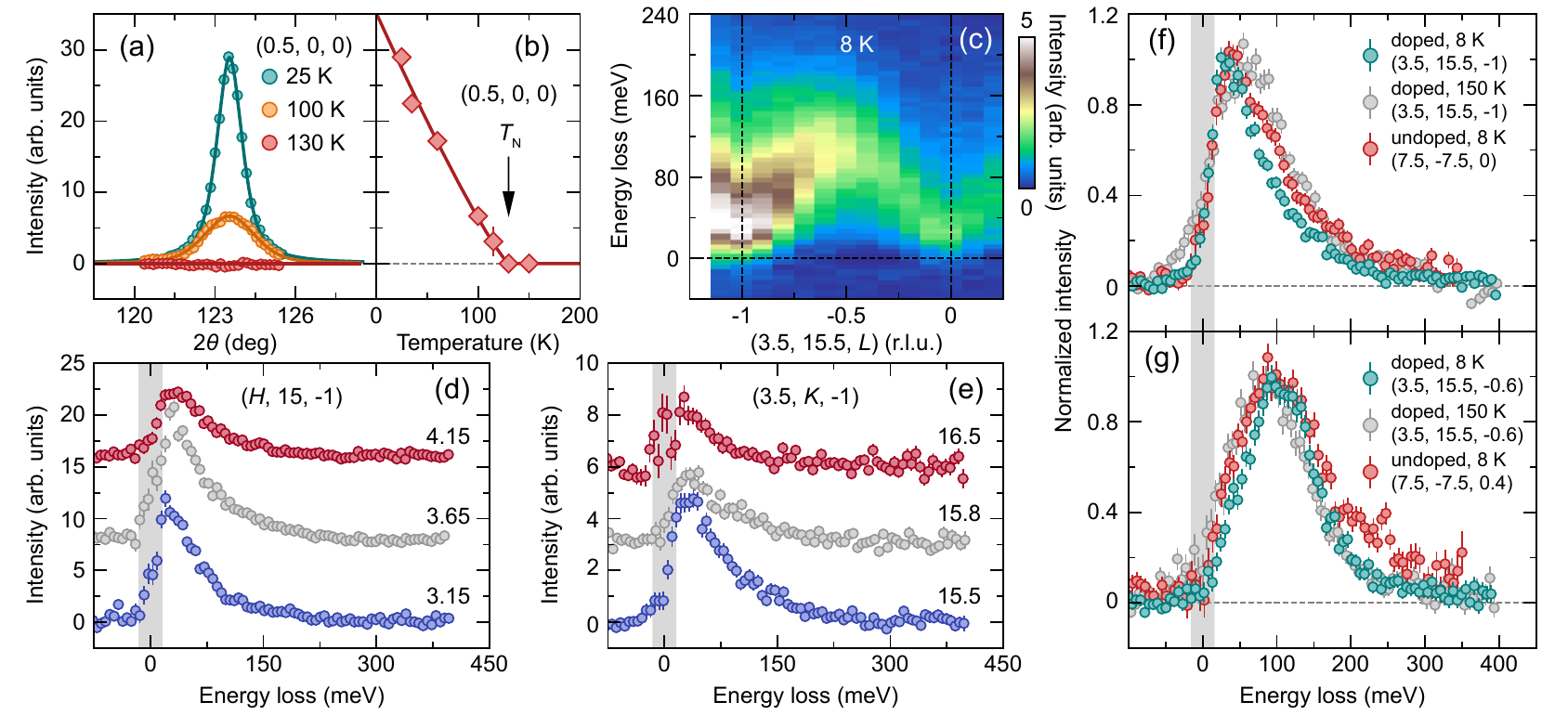}
\caption{Magnetic order and spin dynamics in doped \Sr4310{} ($x=0.02$) and comparison with undoped \Ba4310{}. (a) Background subtracted magnetic Bragg peaks at \Q{}=(0.5, 0, 0), detected at the O $K$-edge with different temperatures in \Sr4310{}. The solid lines are the fitting results using pseudo-Voigt profiles \cite{supp}. (b) Temperature dependence of the fitted peak height. The magnetic transition takes place at \TN{}=130~K. The solid and dashed lines are guides to the eye. (c)-(e) Magnetic excitation spectra of \Sr4310{} at 8~K with different \Q{} along $L$, $H$ and $K$ directions, respectively. (f),~(g) Magnetic excitation spectra at different temperatures for \Ba4310{} and \Sr4310{} samples. The intensities are normalized according to their maximum values. Note that (3.5, 15.5, -1) and (7.5, -7.5, 0) are symmetric regarding the chain direction. The observed phase shift in the spinon spectrum with respect to $L$ arises from zone folding caused by the inter-chain interactions and the sampling of different chains with changes in \Q{} \cite{Bera2017Spinon}. The same applies to (3.5, 15.5, -0.6) and (7.5, -7.5, 0.4). The dashed lines are guides to the eye and the quasi-elastic regime is indicated by the vertical gray bars.}
\label{fig:doped}
\end{figure*}

Figure~\ref{fig:structure}(b) displays a representative Ir $L_3$-edge \gls*{RIXS} spectrum at 11.215~keV composed of a quasi-elastic peak, low-energy magnetic excitations, and high-energy $dd$ excitations of both intra-$t_{2g}$ orbitals and from $t_{2g}$ to $e_g$ orbitals. The overall form of the spectrum is consistent with expectations for Ir $5d^5$ materials \cite{Kim2012Magnetic, Liu2012testing}. We further checked for possible intra-trimer charge disproportionation via bond valence sum analysis and found it to be negligible \cite{supp}. To separate the spectral components we fit the spectra with different functions. The quasi-elastic peak can be represented by a pseudo-Voigt profile with a width dominated by the energy resolution. The low symmetry of \Ba4310{} implies that the magnetic and crystal field excitations are unlikely to have theoretically rigorous analytical forms. We find that phenomenological forms of a damped harmonic oscillator convoluted with the resolution and a manifold of pseudo-Voigts can represent the magnetic and $dd$ excitations respectively, as shown by the fit lines in Fig.~\ref{fig:structure}(b) \cite{supp}. These phenomenological forms for the quasi-elastic line and intra-$t_{2g}$ excitations will be used later to isolate the magnetic scattering which is the main focus of this paper.



We start with the magnetic excitations in undoped \Ba4310{} along different momentum directions. Our sample comes from the same batch studied in Ref.~\cite{Cao2020Quantum} that show no magnetic transition down to 0.2~K. As plotted in Fig.~\ref{fig:undoped}(a),~(b), both the energies and lineshapes are essentially dispersionless in ($H$, 0, 0) and (0, $K$, 0) directions. In contrast, dispersive excitations are revealed along the $L$ direction [Fig.~\ref{fig:undoped}(c)]. Intriguingly, the excitations are very broad along the energy loss axis, distinct from sharp spin waves, rather resembling the magnetic continuum consistent with spinons. A gap-like feature is observed with intensity maximum around 40~meV, consistent with the anisotropy revealed by magnetization \cite{Cao2020Quantum}. Considering the \gls*{1D} character of the dispersion, we propose a spin-1/2 XXZ \gls*{AFM} chain Hamiltonian
\begin{equation}
    \mathcal{H} = J_\text{chain}\sum_{\langle ij \rangle}[S^x_i S^x_j + S^y_i S^y_j + \Delta S^z_i S^z_j]
\end{equation}
where $J_\text{chain}$ is the \gls*{NN} intra-chain interactions, which corresponds to $J_2$ in our case connecting the Ir1 atoms,  $\langle ij \rangle$ denotes bond sums along the chain, and $\Delta$ controls the interaction anisotropy. We calculated the zero temperature two-spinon response with $J_\text{chain}=55$~meV and $\Delta=1.3$ and plot the result in Fig.~\ref{fig:undoped}(c) and Fig.~S5 \cite{Caux2008Two,supp}. This provides a satisfactory description of all the major features of the data, indicating that, despite the \gls*{2D} structure of \Ba4310{}, its magnetic dynamics can be well described by the continuous spinon excitations of \gls*{1D} \gls*{AFM} chains.


To test the fragility of the \gls*{QSL} state, we turn to the isovalently doped \Sr4310{} ($x=0.02$) which has been previously reported to show a magnetic transition below \TN{}=130~K \cite{Cao2020Quantum}. We used O $K$-edge \gls*{REXS} measurements to test for the presence of order and reveal a magnetic Bragg peak at \Q{}=(0.5, 0, 0) [Fig.~\ref{fig:doped}(a),~(b)], which disappears above \TN{}=130~K, consistent with the thermodynamic results \cite{Cao2020Quantum}. Regarding the magnetic excitations, similar with the undoped \Ba4310{}, no dispersion can be distinguished along $H$ and $K$ directions in \Sr4310{}, while clear \Q{} dependence is discovered along $L$ direction at 8~K [Fig.~\ref{fig:doped}(c)-(e)]. A gap-like feature is also observed, consistent with its slight Ising-type anisotropy. However, further analysis of the gap structure, and its correspondence to the heat capacity data, is impeded by the limited energy resolution here. Above \TN{}, the excitations become broader and less dispersive [Fig.~\ref{fig:doped}(f),~(g)]. The flattening of the dispersion is in line with the Ising spin nature.

\begin{figure}
\includegraphics{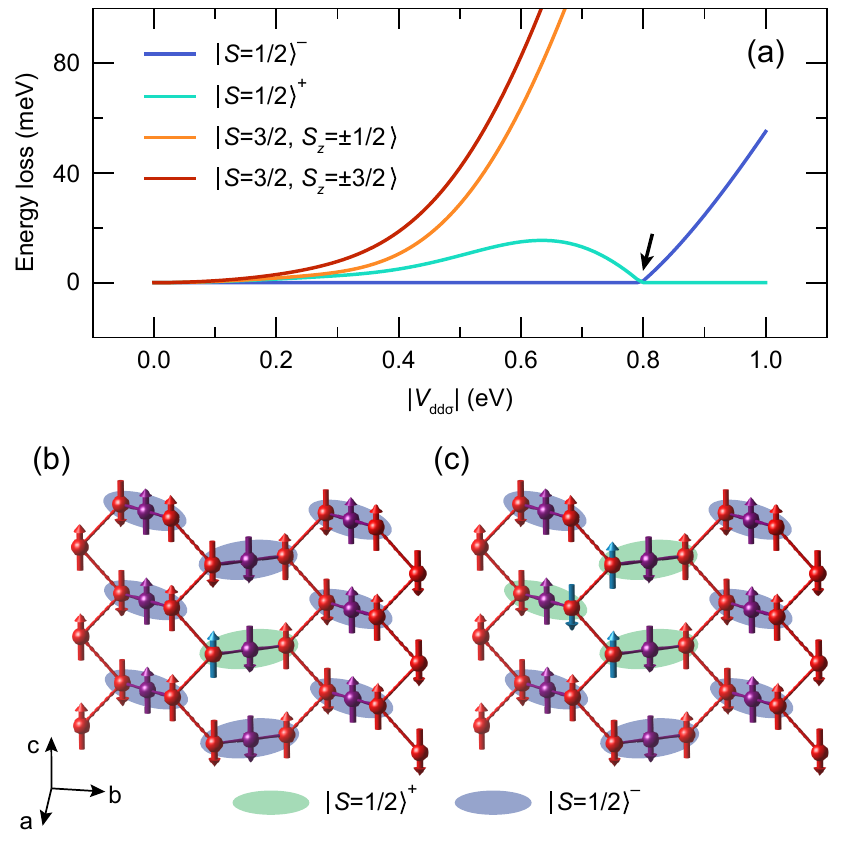}
\caption{Effective chain decoupling. (a) Low-energy spectrum of the three-Ir-site cluster model as a function of inter-site $d$--$d$ hopping \cite{supp}. The black arrow highlights the critical point where the \Sp{} and \Sm{} doublets are degenerate. (b) During the \gls*{RIXS} process, one Ir spin is flipped (cyan arrow), changing the trimer state at low energies and creating two spinons (domain walls). (c) The spinons can propagate in the zigzag chain along $c$ direction by flipping the neighboring spins.}
\label{fig:EDcalcs}
\end{figure}


It has been theoretically proposed that the oxygen-mediated electronic hopping between face-sharing Ir octahedra can be extremely weak under certain circumstances \cite{Kugel2015Spin}, leading to vanishing $J_1$ superexchange parameters, naturally decoupling the zig-zag chains running along $c$ direction in \Ba4310{}. However, this weak coupling model does not take into account the direct hoppings between the neighboring Ir ions which tend to be significant for face-sharing octahedra \cite{Terasaki2016Novel,Okazaki2018Spectroscopic,Ye2018Covalency,Wang2019Direct,Klein2011Antiferromagnetic,Streltsov2012Unconventional,Igarashi2013Xray,Radtke2013Magnetism,Igarashi2015Effects,Sannigrahi2021Orbital}. Indeed, the Bonner-Fisher peak in the susceptibility expected for \gls*{1D} systems \cite{Johnston2000Thermodynamics,Savina2011Magnetic}, is absent in \Ba4310{} \cite{Cao2020Quantum}. Furthermore, compared with undoped \Ba4310{}, the excitations in magnetically ordered \Sr4310{} have similarly shaped dispersion but are somewhat sharper in energy [Fig.~\ref{fig:doped}(f),~(g)]. This behavior differs from that expected for the weak coupling model, where the weak inter-chain coupling leads to multiple bound triplon states in addition to the spinon continuum \cite{Kohno2007Spinons,Coldea2010Quantum,Bera2017Spinon,Mena2020Ising}. Such inconsistency leads us to another explanation, a strong coupling model \cite{Yin2020Frustration,Weichselbaum2021Dimerization}, where the intra-trimer \gls*{AFM} interactions $J_1$ and $J_3$ are substantial and compete with each other, effectively decoupling the chains. 


To understand the origin of the frustration between $J_1$ and $J_3$, we make use of cluster \gls*{ED} calculations utilizing the EDRIXS software \cite{EDRIXS}. We represent the trimer units in \Ba4310{} using a cluster model with three Ir sites with Coulomb interactions and inter-atomic $d$-$d$ hopping explicitly taken into account through Slater-Koster parameterization \cite{supp}. For each Ir site with a nominal $5d^5$ configuration, the ground state is an $S_{\mathrm{eff}}=1/2$ doublet. These $S_{\mathrm{eff}}=1/2$ states recombine in the cluster to yield four Kramers doublets: an antisymmetric doublet \Sm{}, a symmetric doublet \Sp{}, and a high-spin multiplet $|S=3/2\rangle$ which can further split into $|S=3/2, S_z=\pm 1/2\rangle$ and $|S=3/2, S_z=\pm 3/2\rangle$ in the presence of anisotropic interactions \cite{supp}. As shown in Fig.~\ref{fig:EDcalcs}(a), with increasing inter-Ir-site hopping, the energies of $|S=3/2, S_z=\pm 1/2\rangle$ and $|S=3/2, S_z=\pm 3/2\rangle$ increase monotonically. The fact that the $|S=3/2, S_z=\pm 3/2\rangle$ doublet lies at higher energy indicates Ising-like anisotropy for the intra-trimer interactions. In contrast, \Sm{} and \Sp{} show strong competition and the ground state switches between these two doublets at a critical point of $V_{dd\sigma}\approx0.8$~eV.

The presence of the critical point directly leads to the effective decoupling of the spin chains. Assuming an \Sm{} ground state for all the trimers (the real ground state is a superposition of \Sp{} and \Sm{} at the critical point), the \gls*{RIXS} process leads to a single spin flip at one of the Ir sites [Fig.~\ref{fig:EDcalcs}(b)], which at low energies turns the corresponding trimer from \Sm{} to \Sp{}, leaving two spinons (domain walls) in the zigzag spin chain. As the spinons propagate along the chain, the neighboring spins are flipped, switching the trimers between \Sm{} and \Sp{} [Fig.~\ref{fig:EDcalcs}(c)]. At the critical point, as the \Sm{} and \Sp{} doublets are degenerate, such events have no energy cost so that the spinons are deconfined and can propagate freely. Thus, the properties of undoped \Ba4310{} are naturally explained if it sits right at the critical point, resulting in emergent \gls*{1D} behavior and continuous excitations observed in \gls*{RIXS} spectra. The delicate balance between competing interactions can be disrupted easily. In \Sr4310{}, the isovalent Sr doping leads to small structural changes and drives the system slightly away from the critical point so that the spinons become confined since the energy cost to switch between \Sm{} and \Sp{} doublets is no longer zero. Such behavior differs from the conventional doping effect where the dopants could enhance disorder and randomness, which tends to suppress magnetic order and promote glassy physics \cite{Mydosh1993Spin}. It should be noted that the Ising-like anisotropy in \Sr4310{} makes it free from the constraints of Mermin-Wagner theorem which requires a continuous symmetry. Thus, magnetic order does not necessarily imply finite inter-layer coupling, although our diffraction measurements show $H$-axis correlations proving that inter-layer coupling is present. A recent \gls*{REXS} experiment reports a 25~K magnetic transition in the undoped \Ba4310{} with a similar propagation vector as we found in the doped \Sr4310{} \cite{Chen2021Structural}. The origin of this discrepancy is unclear, but it could indicate enhanced disorder.


The strong coupling model puts \Ba4310{} in between the \gls*{1D} spin chains and \gls*{2D} \glspl*{QSL}. Although its magnetic dynamics behave as \gls*{1D} spinon excitations, the underlying magnetism is mostly \gls*{2D}. Consequently, some of the characteristic features for a \gls*{1D} system are missing such as the Bonner-Fisher peak in susceptibility and bound state upon ordering. The \gls*{1D} behavior is caused by the emergent dimensional reduction due to magnetic frustration, which has also been reported in other materials \cite{Okuma2021Dimentional}. This raises an interesting question of whether the properties relevant to \gls*{2D} \glspl*{QSL} are preserved in this case, which is yet to be explored. Measurements of quantum entanglement, such as entanglement witnesses, could prove useful in this regard \cite{Scheie2021Witnessing}.


In summary, we use \gls*{RIXS} to show that emergent \gls*{1D} spinon excitations can arise from the \gls*{2D} magnetism in \Ba4310{} due to the frustrated inter-chain (intra-trimer) interactions. The highly suppressed magnetic order can be easily recovered by disturbing the subtle balance of the frustration, confining the spinons into magnons. Although prior work has speculated that \Ba4310{} could be either a Luttinger liquid \gls*{QSL} or a 2D \gls*{QSL} such as a spinon Fermi surface states \cite{Cao2020Quantum}, the data here are the first evidence of a 1D spinon continuum in \Ba4310{}. These results indicate that, instead of forming an isotropic \gls*{QSL} state, magnetic frustration can effectively reduce the system dimension, suppressing the magnetic order and realizing deconfined spinons in a unique way.

\begin{acknowledgments}
We thank Emil Bozin and Kemp Plumb for insightful conversations. Work at Brookhaven National Laboratory was supported by the U.S. Department of Energy, Office of Science, Office of Basic Energy Sciences. G.C.\ acknowledges NSF support via grant DMR 1903888. This research used resources of the Advanced Photon Source, a U.S. Department of Energy (DOE) Office of Science User Facility at Argonne National Laboratory and is based on research supported by the U.S. DOE Office of Science-Basic Energy Sciences, under Contract No. DE-AC02-06CH11357. This research used resources at the Coherent Soft X-Ray and In Situ and Resonant Hard X-ray Studies beamline  of the National Synchrotron Light Source II, a U.S. Department of Energy (DOE) Office of Science User Facility operated for the DOE Office of Science by Brookhaven National Laboratory under Contract No. DE-SC0012704.

\end{acknowledgments}

\bibliography{refs}
\end{document}


\title{Supplemental Material: Emergence of spinons in layered trimer iridate Ba$_4$Ir$_3$O$_{10}$}

\renewcommand{\thepage}{S\arabic{page}} 
\renewcommand{\thesection}{S\arabic{section}}  
\renewcommand{\thetable}{S\arabic{table}}  
\renewcommand{\thefigure}{S\arabic{figure}}

\date{\today}

\maketitle

This document provides methods information, the bond valence sum analysis, additional details of the X-ray measurements, and description of the lineshape fitting.

\section{Methods}

Single crystals of \Ba4310{} were grown using the flux method as described in Ref.~\cite{Cao2020Quantum}. The \gls*{RIXS} experiments were performed at the 27-ID-B station of the Advanced Photon Source at Argonne National Laboratory. All \gls*{RIXS} data were collected at the Ir $L_3$-edge with a horizontal scattering plane and $\pi$ polarization. Wavevectors throughout the manuscript are defined using standard reciprocal lattice units (r.l.u.) notation as \Q{}$=H\mathbf{a}^{\ast}+K\mathbf{b}^{\ast}+L\mathbf{c}^{\ast}$ based on lattice constants $a=$7.2545~\AA, $b=$13.192~\AA, $c=$5.7737~\AA, $\alpha=\gamma=90^{\circ}$, $\beta=113.513^{\circ}$ and space group $P2_1/c$ (No.~14). The energy resolution was around 32~meV. The \gls*{REXS} measurements were carried out at the Coherent Soft X-Ray (CSX) 23-ID-1 beamline at the National Synchrotron Light Source II with x-ray energy tuned to the O $K$-edge, $\pi$ x-ray polarization and a vertical scattering geometry. Due to the \gls*{SOC} of Ir and hybridization between Ir and O, O $K$-edge \gls*{REXS} is sensitive to magnetic order in iridates \cite{Liu2015probing}. Meanwhile, the small penetration depth and long wavelength of O $K$-edge x-rays help suppress multi scattering events which can present a challenge in Ir $L_3$-edge \gls*{REXS} experiments.

\section{Bond valence sum analysis}

The bond valence sum is a widely utilized means to estimate the oxidation states of atoms in a compound. Here we use it to test the possibility of intra-trimer charge disproportionation in \Ba4310{}. The valence $V$ of an atom can be evaluated by summing up the individual bond valences $v_i$ surrounding the atom, which have a simple relationship with the bond lengths \cite{Brown2006Chemical}:
\begin{equation}
    V=\sum_i{v_i}=\sum_i{\exp(\frac{R_o-R_i}{B})}
\end{equation}
where $B=0.37$~\AA{} is an empirical constant, $R_i$ is the bond length in the target material, and $R_0$ is the reference bond length for an element of exact valence, for which we use tabulated values of 1.87~\AA{}, derived for Ir$^{4+}$-O$^{2-}$ bonds \cite{VBSparams}. We use the \Ba4310{} structure reported in Ref.~\cite{Cao2020Quantum}. The corresponding valence is 4.16+ and 3.87+ for Ir1 and Ir2, respectively, indicating that charge disproportionation is small and of minimal importance for the physics discuss here. It is certainly far too small to generate an effectively spinless central iridium atom, as has been proposed for Ru based material Ba$_4$Ru$_3$O$_{10}$ \cite{Streltsov2012Unconventional}. We note that inter-trimer disproportionation is forbidden since the trimers are symmetrically equivalent in \Ba4310{}.

\section{Fitting of the REXS data}

\begin{figure*}
\includegraphics{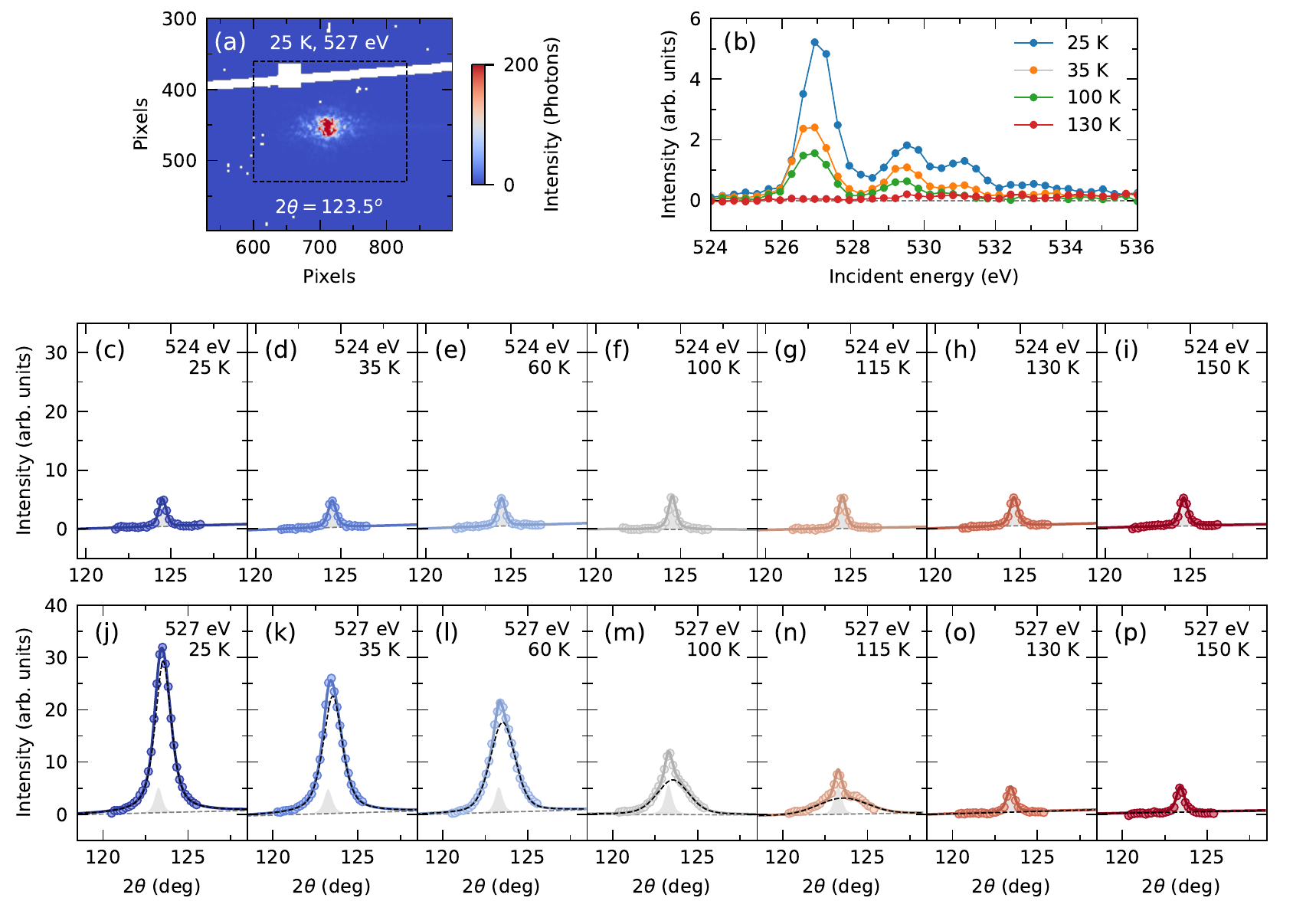}
\caption{Fitting of the \gls*{REXS} data at O $K$-edge. (a) A representative detector image of the superlattice peak at \Q{}=(0.5, 0, 0). The dashed lines envelop the region of interest in which the pixel readings are summed up to extract the peak intensity. The white pixels originate from the beamstop/detector errors that have been removed from the data. (b) Incident energy dependence of the superlattice peak intensity at different temperatures. The small energy shift ($<0.3$~eV) between the energy scans of different temperature is due to the photon energy change associated with the monochromator that is sensitive to thermal perturbations. (c)-(p) A series of theta-2theta scans with different incident energy and temperature. The circles are the data and the solid lines are the fitting results which can be further broken into several components including a sloped background (gray dashed lines), a pseudo-Voigt function presenting the second harmonic of the Bragg peak (gray shaded area) and another pseudo-Voigt function outlining the magnetic contributions (black dashed lines).}
\label{fig:REXS_raw}
\end{figure*}

For the \gls*{REXS} measurements, we used a piece of \Sr4310{} ($x=$0.02) single crystal with a surface normal of [$H$, 0, 0]. We put the [$H$, 0, 0] and [0, 0, $L$] directions in the vertical scattering plane and used $\pi$ incident x-ray polarization. The Coherent Soft X-Ray (CSX) 23-ID-1 beamline at the National Synchrotron Light Source II has a fast CCD detector with $30 \times 30$~\um$^2$ pixels placed 340~mm from the sample. At 25~K, a well-defined peak emerged in the fast CCD image [Fig.~\ref{fig:REXS_raw}(a)], corresponding to the superlattice peak at \Q{}=(0.5, 0, 0). The speckle pattern arises from the interference of the coherent x-ray beam scattered by different magnetic domains \cite{Shen2021charge}. The peak intensity is extracted by summing up the pixel readings in the region of interest that is marked by the dashed rectangle in Fig.~\ref{fig:REXS_raw}(a). Figure~\ref{fig:REXS_raw}(b) shows the incident energy dependence of the peak intensity at fixed \Q{}=(0.5, 0, 0) in which resonant behavior can be observed. The multiple resonant peaks may be caused by the symmetry-distinct oxygen sites that have different binding energies. By comparing the theta-2theta scans at the non-resonant (524~eV) and resonant (527~eV) energies [Fig.~\ref{fig:REXS_raw}(c)-(p)], we find two different contributions to the signals. One is a tiny peak that is independent of incident energy or temperature (gray shaded areas), which comes from the second harmonic of the (1, 0, 0) Bragg peak. Another contribution shows clear resonant behavior (black dashed lines), indicating a magnetic origin. It weakens with increasing temperature and disappears above 130~K, consistent with previously reported magnetic transition \cite{Cao2020Quantum}. The pure magnetic contribution can be isolated by fitting the theta-2theta scan with a sloped background and two pseudo-Voigt profiles, the results of which are shown in Fig.~3(a),~(b) of the main text.

\section{Fitting of the RIXS spectra}

\begin{figure*}
\includegraphics{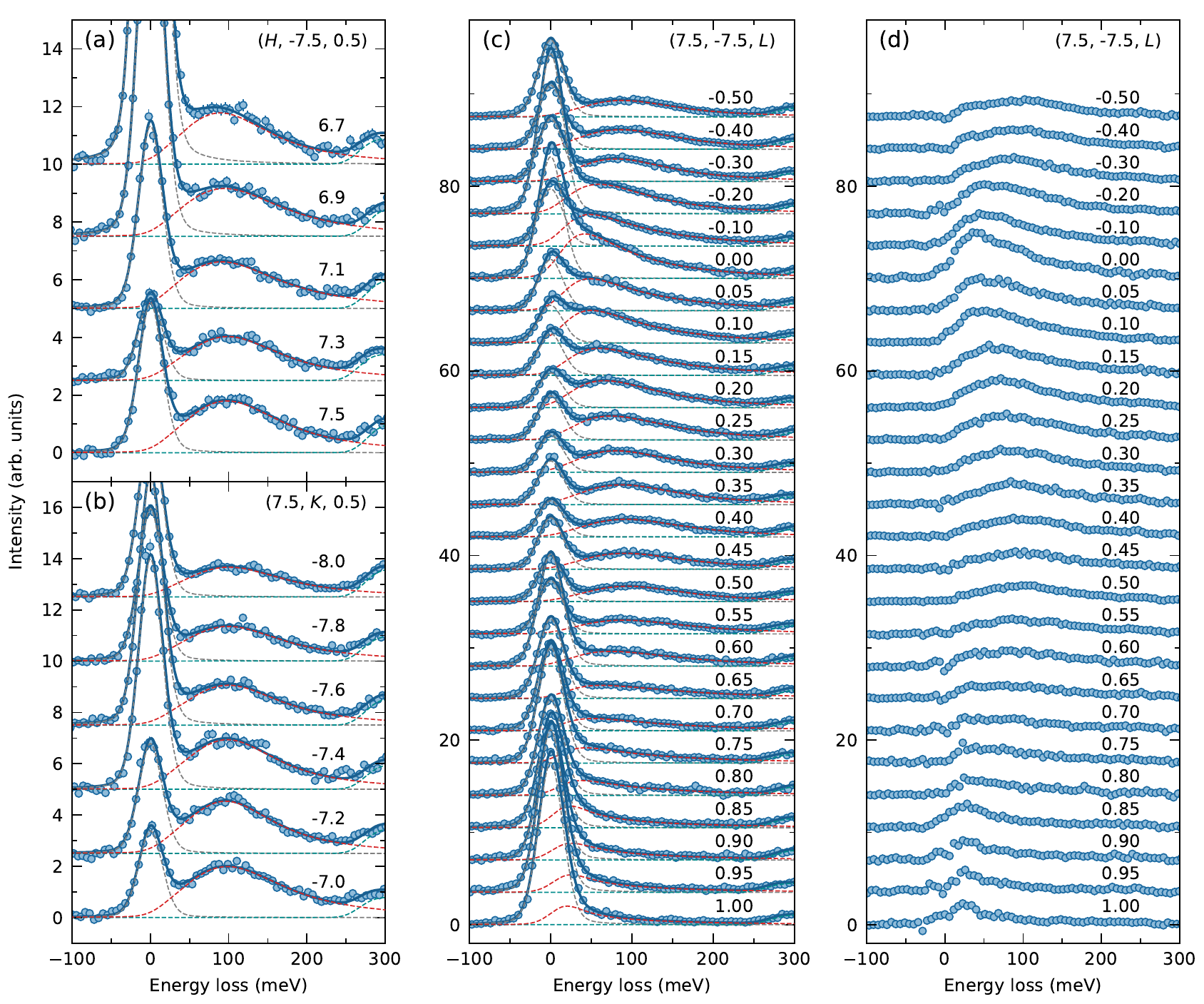}
\caption{Fitting of the Ir $L_3$-edge \gls*{RIXS} data in undoped \Ba4310{} collected at 8~K. (a)-(c) Background subtracted \gls*{RIXS} spectra at different \Q{} points. The circles present the data and the dashed lines of different colors are fittings of different components, the summation of which leads to the solid lines. (d) \gls*{RIXS} spectra with the quasi-elastic line and $dd$-excitations subtracted. The same set of data produces Fig.~2(c) in the main text.}
\label{fig:undoped_raw}
\end{figure*}

\begin{figure*}
\includegraphics{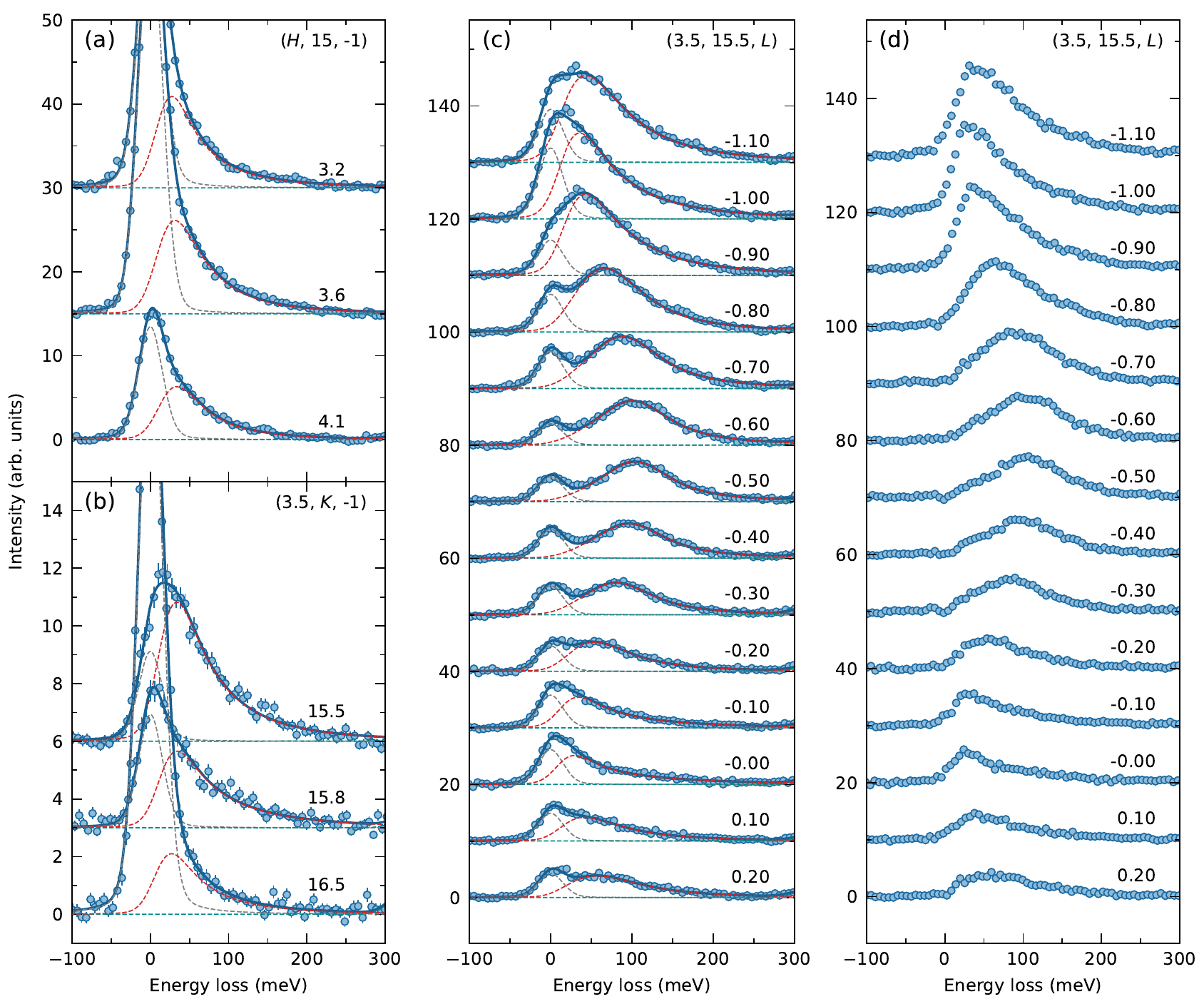}
\caption{Fitting of the Ir $L_3$-edge \gls*{RIXS} data in doped \Sr4310{} ($x=0.02$) collected at 8~K. (a)-(c) Background subtracted \gls*{RIXS} spectra along with fitting results presented by dashed lines of different colors. (d) \gls*{RIXS} spectra with the quasi-elastic line and $dd$-excitations subtracted. The same set of data produces Fig.~3(c) in the main text.}
\label{fig:doped_raw}
\end{figure*}

\begin{figure*}
\includegraphics{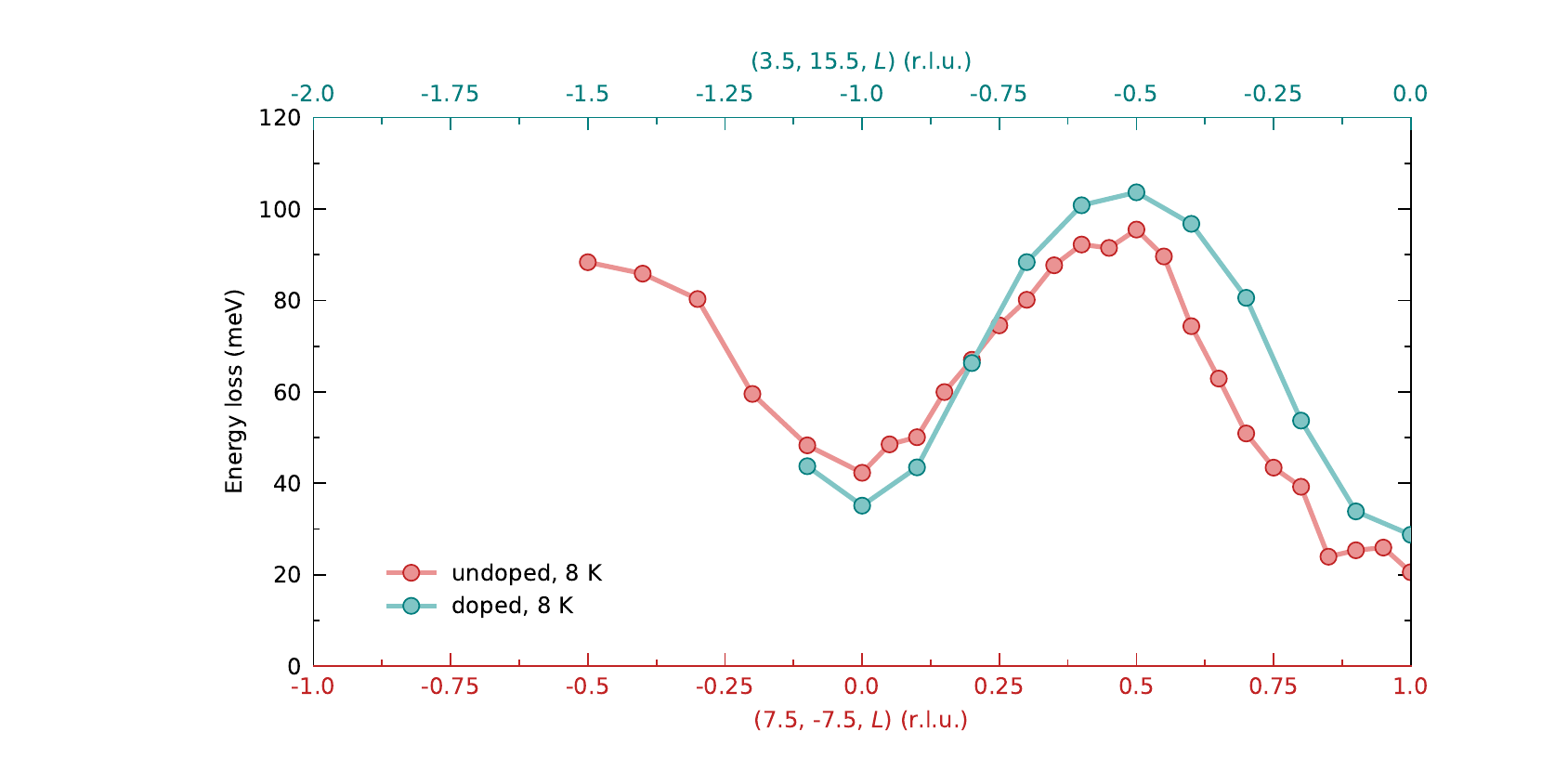}
\caption{Energy positions of magnetic excitation maxima as a function of $L$ for undoped \Ba4310{} and doped \Sr4310{} ($x=0.02$).}
\label{fig:disp}
\end{figure*}

As shown in Fig.~1(b), a representative \gls*{RIXS} spectrum is composed of a quasi-elastic line, low-energy magnetic excitations and high-energy $dd$-excitations. A constant background has been subtracted for all the spectra using the energy-gain side. The quasi-elastic peak includes the elastic line and potential phonon excitations which turn out to be negligible. Thus, we fit the quasi-elastic peak with a pseudo-Voigt profile that is primarily determined by the energy resolution. The lineshape of the magnetic excitations can be adequately captured phenomenologically by a damped harmonic oscillator convoluted with the resolution. In Fig.~1(b) of the main text, the $dd$ excitations are fitted with multiple pseudo-Voigt functions. For Fig.~2 and Fig.~3, the $dd$ excitations are not fully covered during the measurements, so we use an error function to account for the tail of $dd$-excitations for simplicity. The fitting results are plotted in Fig.~\ref{fig:undoped_raw}(a)-(c) and Fig.~\ref{fig:doped_raw}(a)-(c). After subtracting the quasi-elastic line and $dd$-excitations, the magnetic excitations are well isolated, which are presented in Fig.~2(a)-(c), Fig.~3(c)-(e), Fig.~\ref{fig:undoped_raw}(d), and Fig.~\ref{fig:doped_raw}(d). The extracted dispersion along the $L$ direction is plotted in Fig.~\ref{fig:disp}, in which the energy positions of intensity maxima from fits are used since the harmonic oscillator is overdamped near the zone center.

\section{Details of the cluster exact diagonalization calculations}

\begin{figure}
\includegraphics{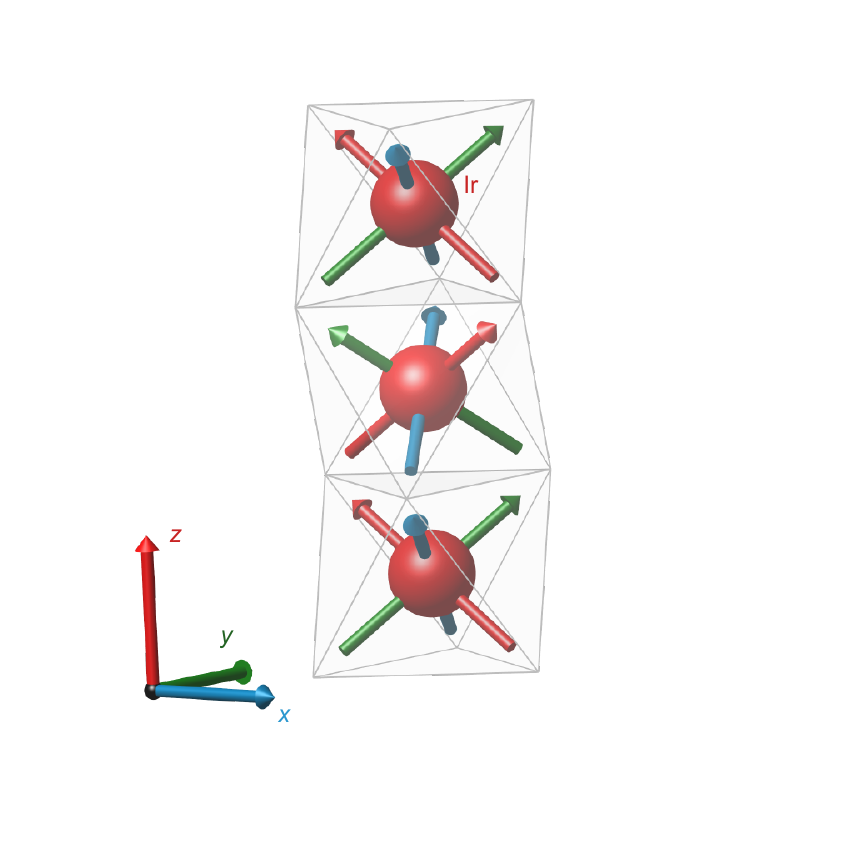}
\caption{Illustration of the cluster model used in the ED calculations. Both the global Cartesian coordinates defined in the trigonal notation and the local coordinates defined in the octehedral notation are presented with the same color coding.}
\label{fig:ED_model}
\end{figure}

To help understand the strong coupling model, we perform cluster \gls*{ED} calculations using the EDRIXS software \cite{EDRIXS}. A three-Ir-site cluster is constructed as shown in Fig.~\ref{fig:ED_model}. We explicitly include the onsite Coulomb interactions and spin orbital couplings in the Hamiltonian. Based on the short Ir-Ir bond distances and the close-to-right-angle Ir-O-Ir bonds, the leading interactions between the Ir atoms are expected to be direct hopping, so we proceed based on the same approach as that used in Ref.~\onlinecite{Kugel2015Spin} and implement inter-atomic $d$-$d$ hopping integrals between the $5d$ orbitals of the neighbored Ir sites using the Slater-Koster method. The interactions are first implemented in the trigonal notation (global coordinates) and are transformed into octahedral notation (local coordinates) subsequently. Regarding the orbital basis, we consider $t^{15}_{2g}e^{0}_{g}$ and $t^{14}_{2g}e^{1}_{g}$ configurations as the double $e_g$ occupancy is minimal. The full set of parameters used for the \gls*{ED} calculations is listed in Table~\ref{table:allparams}.

As mentioned in the main text, in the absence of inter-Ir-site hopping, the ground state for each Ir site with a nominal $5d^5$ configuration is an $S_{\mathrm{eff}}=1/2$ doublet. With finite inter-Ir-site hopping, the three doublets from the three Ir sites will recombined and split into several multiplets. For the antisymmetric doublet \Sm{}, the two Ir sites at the edge of the trimer form a singlet and the spin of the middle Ir site remains free. In the symmetric doublet \Sp{}, the two Ir sites at the edge form a triplet which keeps antiparallel to the middle Ir site spin. These two doublets are protected by symmetry. An additional high-spin multiplet $|S=3/2\rangle$ lies at higher energy than these two doublets given the antiferromagnetic interactions, the degeneracy of which is lifted by anisotropic interactions, making it split into $|S=3/2, S_z=\pm 1/2\rangle$ and $|S=3/2, S_z=\pm 3/2\rangle$ doublets. The transitions between \Sm{} and \Sp{} doublets contribute to the magnetic signals below 0.25~eV while the transitions to $|S=3/2\rangle$ multiplets are indicated by the peak at $\sim$0.35~eV in Fig. 1(b). As shown in the main text, the competition among intra-trimer interactions leads to a critical point where the \Sm{} and \Sp{} doublets become degenerate. The exact position of the critical point depends on both the Coulomb interactions and spin orbital coupling.

The presence of the critical point is a consequence of the competition between $J_1$ and $J_3$. Indeed, in a mean-field level, \Sp{} and \Sm{} becomes degenerate when $J_1=J_3$ in the isotropic case \cite{Weichselbaum2021Dimerization}. This form of calculation between the interactions can explain both the emergent 1D behavior in \Ba4310{} and the fragility of the \gls*{QSL} state in \Sr4310{}. Upon doping, the system is slightly driven away from the critical point, making the spinons more confined. Considering that the changes of lattice parameters and atomic coordinates on doping are minimal \cite{Cao2020Quantum}, the modification of the exchange interactions is expected to be subtle which can be driven by either the change of inter-site hopping or the screening of Coulomb interactions caused by the surrounding oxygens. Thus, the intra-chain interactions, both $J_{\mathrm{chain}}$ and $\Delta$, would be largely preserved, which is consistent with the observation that the overall dispersion along $L$ direction remains quite similar in the doped sample [Fig.~2(c) and Fig.~3(c) in the main text].

\begin{table*}
\caption{Full list of parameters used for the \gls*{ED} calculations. Here, we only consider the cubic crystal field splitting and fix $V_{dd\pi}=-2/3V_{dd\sigma}$, $V_{dd\delta}=1/3V_{dd\sigma}$ \cite{Anderson1978Electronic}. $F^0_{dd}$, $F^0_{dd}$ and $F^0_{dd}$ are Slater integrals for the Ir $5d$ orbitals and $\zeta$ is the spin orbital coupling strength. All parameters are in units of eV.}
\begin{ruledtabular}
\begin{tabular}{ccccc}
$10D_q$ & $F^0_{dd}$ & $F^2_{dd}$ & $F^4_{dd}$ & $\zeta$ \\ 
\hline
3.5 & 2.66 & 2.58 & 1.62 & 0.33 \\
\end{tabular}
\end{ruledtabular}
\label{table:allparams}
\end{table*}

\section{Additional calculations of the spinon continuum}

\begin{figure}
\includegraphics{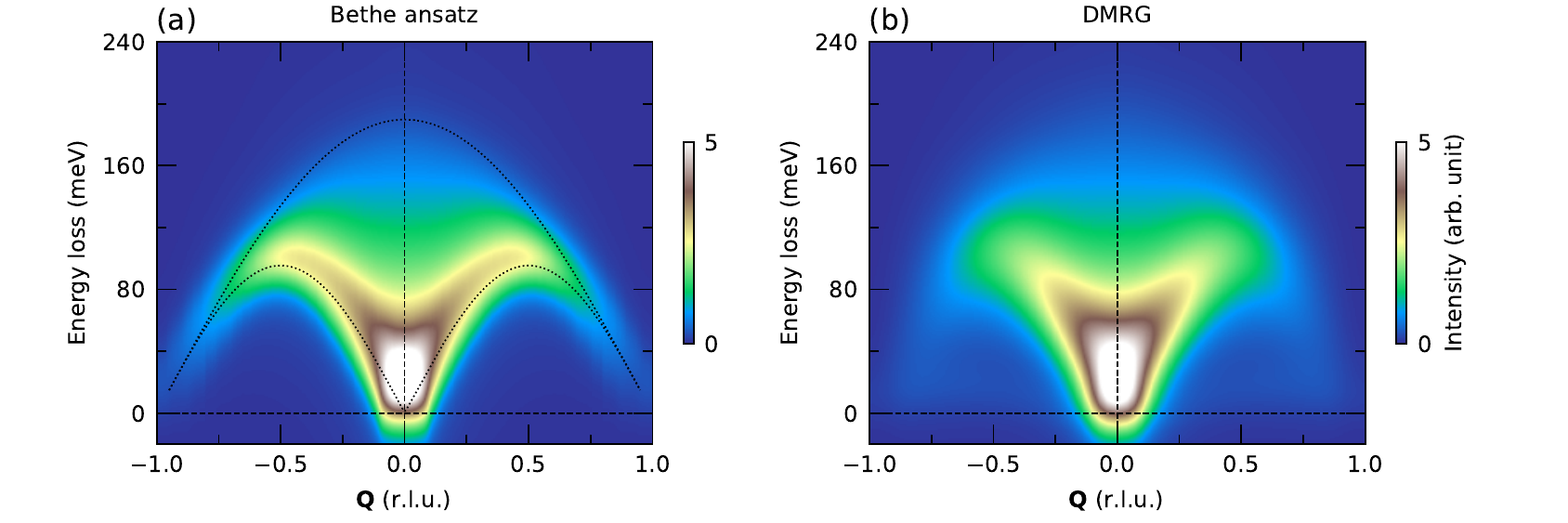}
\caption{Calculated spinon excitations along the chain direction. (a) Two-spinon dynamical structure factor of an antiferromagnetic spin-1/2 chain calculated using Bethe ansatz with $J_{\mathrm{chain}}=55$~meV and $\Delta=1.3$. The dotted lines are the calculated excitation boundaries without broadening. (b) Dynamical structure factor calculated using DMRG approach as described in Ref.~\onlinecite{Weichselbaum2021Dimerization} with $J_1^{xx}=181.5$, $J_2^{xx}=45.375$, $J_3^{xx}=195.1125$~meV and $\Delta=2$. Only the contributions from the Ir atoms within the chains are presented. The spectra in both panels are convoluted with instrument resolution function and the dashed lines are guides to the eye.}
\label{fig:spinon}
\end{figure}

In Fig.~2(c) of the main text, we show the excitation boundaries of the spinon continuum. Here, we present more results about the spinon excitation calculations. The calculated spinon continuum spectrum is shown in Fig.~\ref{fig:spinon}(a), using the same parameters as in the main text. Our effective model further captures the energy-width and boundaries of the continuum, but only approximates the detailed lineshape. This could arise from further sub-leading exchange interactions in the full magnetic Hamiltonian for \Ba4310{}, such as the off-diagonal terms, which could come into play given the strong \gls*{SOC} and low structural symmetry of \Ba4310{}. Note that the contributions from higher order four-spinon excitations fade rapidly with increasing $\Delta$ \cite{Perez2020exact}.


Moreover, the spinon continuum we calculated here only involves the zigzag chains formed by the Ir sites at the edge of the trimers. The presence of the middle Ir sites is expected to provide extra correlations, making the lineshape deviate from a simple picture of an antiferromagnetic spin chain. To test with a more realistic model and further explore the effect of exchange anisotropy on spinon excitations, we performed additional \gls*{DMRG} calculations within the strong coupling regime using the same model as described in Ref.~\onlinecite{Weichselbaum2021Dimerization}, which demonstrates the intra-trimer frustration in the isotropic limit. With $\Delta>1$, since the transformation of $S_z\rightarrow -S_z$ leaves the Hamiltonian invariant, the low-energy multiplet still represents two $S=1/2$ multiplets, \Sm{} and \Sp{}, that cross in a linear fashion when tuning $J_1$ across $J_3$. Consequently, the calculated spectrum is qualitatively the same as in the isotropic case. Figure~\ref{fig:spinon}(b) shows the results with $\Delta=2$, which is larger than the value determined from \gls*{RIXS} data in order to enhance the spin gap. It turns out that the spin gap is rather small despite a very sizeable $\Delta$. The calculated spectrum shows robustly a spinon continuum but with a modified lineshape.

Overall, the current minimal model turns out to be a good description of the most important interactions in \Ba4310{}.

\clearpage
\bibliography{refs}